\documentstyle[multicol,aps,epsf]{revtex}

\begin{document}
\title{Competition of Mesoscales and Crossover to Tricriticality in Polymer
Solutions}
\author{M. A. Anisimov, A. F. Kostko, and J. V. Sengers}
\address{Department of Chemical Engineering \\
and Institute for Physical Science and\\
Technology, University of Maryland, College Park, MD 20742}
\date{\today}
\maketitle

\begin{abstract}
We show that the approach to asymptotic fluctuation-induced critical
behavior in polymer solutions is governed by a competition between a
correlation length diverging at the critical point and an additional
mesoscopic length-scale, the radius of gyration. Accurate light-scattering
experiments on polystyrene solutions in cyclohexane with polymer molecular
weights ranging from 200,000 up to 11.4 million clearly demonstrate a
crossover between two universal regimes: a regime with Ising asymptotic
critical behavior, where the correlation length prevails, and a regime with
tricritical {\it theta}-point behavior determined by a mesoscopic
polymer-chain length.
\end{abstract}

\pacs{PACS: 64.75.+g; 61.25.Hg; 05.70.Jk}

\protect\tightenlines

\begin{multicols}{2}
Close enough to the critical point, the correlation length $\xi $ of the
fluctuations of the order parameter has grown so large that the microscopic
and even the mesoscopic structure of fluids become unimportant: complex
fluids become ''simple''. This feature is known as critical-point
universality \cite{F:82}. Within a universality class, determined by the
nature of the order parameter, properly chosen physical properties of
different systems exhibit the same near-critical behavior. All critical
phase-separation transitions in fluids belong to the 3-dimensional
Ising-model universality class, as the order parameter (associated with
density or/and concentration) is a scalar. However, in practice, the pure
asymptotic regime is often hardly accessible. Even in simple fluids, like
xenon and helium, the physical properties in the critical region show a
tendency to crossover from Ising asymptotic behavior to mean-field behavior 
\cite{GC:81,HR:01}. This crossover depends on the microscopic structure of
the system, namely, on the range of interaction and on a molecular-size
''cutoff''. In simple fluids, crossover to mean-field critical behavior is
never completed within the critical domain (which can be defined roughly as
within 10\% of the critical temperature): the ''cutoff'' length and the
range of interactions are too short. In complex fluids, regardless of the
range of interaction, the role of the cutoff is played by a mesoscopic
characteristic length scale $\xi _{{\rm D}}$ that is associated with a
particular mesoscopic structure \cite{AS:95}. If the cutoff length is
mesoscopic, it can compete with the correlation length $\xi $ within the
critical domain. The temperature at which the correlation length becomes
equal to the structural length can be naturally defined as a crossover
temperature between two regimes, namely, an Ising asymptotic critical regime
and a regime determined by the nature of the mesoscopic structure of the
complex fluid. In some complex fluids, like polymer solutions, it is
possible to tune the structural length-scale and make it very large. If both
lengths, the correlation length of the critical fluctuations associated with
the fluid-fluid separation and the structural correlation length, diverge at
the same point, this point will be a multicritical point. A perfect example
of such a multicritical phenomenon appears in a polymer solution near the 
{\it theta} point. The {\it theta} point is the point of limiting (infinite
molecular weight of polymer) of phase separation in an infinitely dilute
solution \cite{Fl:53}. It is well known that the {\it theta}-point is a
''symmetrical tricritical point''\cite{DG:79}. Both the radius of gyration $%
R_{{\rm g}}$ (assumed to be proportional to $\xi _{{\rm D}}$) of the polymer
molecule and the correlation length of the concentration fluctuations
diverge at the {\it theta} point, leading to tricriticality. 
Phenomenologically, a tricritical point emerges 
because of a coupling between two order
parameters, one scalar and one vector-like, which results in a change in the
order of the phase transition: a second-order transition associated with the
structural order parameter becomes first-order and thus is accompanied by
phase separation \cite{KS:84}. At a tricritical point, two universality
classes meet each other, making tricritical behavior (in three dimensions)
almost mean-field like. Thus, the radius of gyration serves as a ''screening
length'' for the critical fluctuations of concentration. By tuning the
radius of gyration (probing different molecular weights) and the correlation
length of critical fluctuations (changing the temperature distance from the
liquid-liquid critical point), one can probe crossover from Ising asymptotic
behavior to mean-field-like tricritical {\it theta}-point behavior. Such a
crossover was first detected from analyses of the osmotic susceptibility
obtained in a neutron scattering experiment \cite{MH:97} and of the shape of
polymer solution coexistence curves \cite{AG:01}. Unfortunately,
experimental data analyzed so far were 
not close enough to the {\it theta} point
to unambiguously separate crossover to tricriticality from nonasymptotic
regular effects \cite{PS:99}.

To observe the competition of two mesoscales in polymer solutions and
convincingly prove crossover to {\it theta}-point tricriticality, one needs to
tune both the radius of gyration and the correlation length of the critical
fluctuations over a large range, that is to probe as high molecular weights
as possible, and to measure both the critical susceptibility and the
correlation length with an accuracy of the order of 1\%. It is a challenging
experimental task and even most recent light-scattering experiments in
polymer solutions \cite{SS:99,RW:99} did not resolve crossover to mean-field
{\it theta}-point behavior.

In this communication we report accurate light-scattering experiments
performed for polystyrene solutions in cyclohexane with polymer molecular
weight, $M_{{\rm w}}$, ranging from 200,000 to 11.4 million. The data
clearly and unambiguously confirm the physical nature of crossover to
{\it theta}-point tricriticality as described above. Moreover, the crossover
behavior of the susceptibility and correlation length is in excellent
agreement with the theory of crossover critical phenomena, also developed at
the University of Maryland (see\cite{AT:92,AS:01} and bibliography there).
We have also attempted to solve one of the \ most subtle problems of
tricriticality, namely, experimental detection of so-called logarithmic
corrections to mean-field tricritical behavior \cite{D:82,HS:99}. Such
corrections have a universal nature as they are predicted theoretically to
exist near all kinds of tricritical points \cite{LS:84}.

Experimental technique, sample preparation, and experimental procedure have
been described in detail elsewhere \cite{JG:01}. Five polystyrene samples\
(obtained from Polymer Laboratories Inc.) with a polydispersity index 1.02 ($%
M_{{\rm w}}$=1.96$\cdot $10$^{5}$), 1.06 ($M_{{\rm w}}$=1.12$\cdot $10$^{6}$%
), 1.04 ($M_{{\rm w}}$=1.95$\cdot $10$^{6}$), 1.05 ($M_{{\rm w}}$=3.95$\cdot 
$10$^{6}$), and 1.09 ($M_{{\rm w}}$ = 11.4$\cdot $10$^{7})$ have been
investigated. Two identical He-Ne lasers and a receiving photomultiplier
system have been aligned at two fixed scattering angles, 30${{}^\circ}$
and 150${{}^\circ}$. Small parts of the incident-beam intensity 
of both lasers, directed by
means of beam splitters and optical guides to the photomultiplier, served as
calibration intensities. Our measurement procedure allows elimination of the
influence of any laser power drift, as well as of slow fluctuations in the
sensitivity of the photomultiplier. A square optical cell with an optical
path of 2 mm is placed in a two-stage thermostat. This system allows
stabilization of the temperature to within 0.5 mK over a few days. We
determined the critical temperature $T_{{\rm c}}$ (more precisely, the
temperature of phase separation) by monitoring the intensity of the
transmitted beam and the scattering intensity while the sample is cooled
from above the critical temperature in steps of 2-3 mK. The critical
composition $\phi _{{\rm c}}$ was checked by equality of volumes of the
coexisting phases and established with an accuracy of at least 3-5\%.
Turbidity measurements were made enabling us to apply the corrections due to
turbidity loss and to multiple scattering evaluated by a Monte-Carlo
simulation \cite{BC:94,SS:99}. The overall accuracy of the intensity
measurements in the range of $\tau =(T-T_{{\rm c}})/T_{{\rm c}}$ varying
from $10^{-6}$ to $10^{-1}$ is estimated to be about 1-2\%. Data at $\tau
<10^{-5}$ and $\tau >$ 6$\cdot $10$^{-2}$ were not included in the analysis
as they became strongly affected by uncertainty in the critical composition
and by polydispersity (close to $T_{{\rm c}}$), and by background scattering
(far away from $T_{{\rm c}}).$

The corrected scattering intensity, $I$, was fitted to the following
expression

\begin{equation}
I=I_{0}\chi G\left( q\xi \right) +I_{{\rm b}}\text{ .}  \label{Intensity}
\end{equation}

Here $\chi $ is the osmotic susceptibility, $I_{{\rm b}}$ is a background
intensity, $I_{0}$ is an instrumental constant, $q=4\pi n\lambda
_{0}^{-1}\sin (\theta /2)$ is the scattering wave number ($n$ the refractive
index, $\lambda _{0}$ the wave length of the incident light , $\theta $ the
scattering angle), and $G(q\xi )$ is the spatial correlation function taken
in the Fisher-Burford approximation \cite{FB:67},

\begin{equation}
G\left( q\xi \right) =\frac{[1+0.084^{2}(q\xi )^{2}]^{\eta /2}}{1+(q\xi
)^{2}(1+\frac{\eta }{2}0.084^{2})}  \label{corrfunc}
\end{equation}
with $\eta =0.033,$ a universal critical exponent.

The osmotic susceptibility and the correlation length were represented by
the following crossover expressions \cite{MH:97} taken for the particular
case of the normalized coupling constant $\overline{u}=1$: 

\begin{equation}
\chi ^{-1}=a_{0}\tau Y^{(\gamma -1)/\Delta _{{\rm S}}}[1+\frac{u^{\ast }\nu
z^{2}}{2(z^{2}+2\nu )}]\text{ ,}  \label{suscept}
\end{equation}

\begin{equation}
\xi =\overline{\xi }_{0}\tau ^{-1/2}Y^{(1-2\nu )/2\Delta _{{\rm S}}}\text{ ,}
\label{corrlength}
\end{equation}
where $Y$ is a crossover function of a single argument, $z=\xi /\xi _{{\rm D}%
},$ defined as

\begin{equation}
Y=(1+z^{2})^{-\Delta _{{\rm S}}/2\nu }\text{ .}  \label{corfunc}
\end{equation}

In Eqs.(3)-(5), $\gamma =1.239,\nu =0.630,$ and $\Delta _{{\rm S}}\cong 0.50$
(we adopted $\Delta _{{\rm S}}=0.51$ \cite{AT:92}) are universal critical
exponents and $u^{\ast }=0.472$ for the 3-dimensional Ising universality
class\cite{F:82,AT:92}; $a_{0}$ and $\overline{\xi }_{0}$ are mean-field
amplitudes of the inverse susceptibility and of the correlation length. In
principle, the crossover function depends on two crossover parameters, $z$
and a normalized coupling constant $\overline{u}$. However, the analysis of
the experimental data has shown that $\overline{u}$ is almost independent of 
$M_{{\rm w}}$ and always close to unity. Since a theoretical analysis \cite
{PS:99} confirms that this parameter is irrelevant for crossover to
{\it theta}-point tricriticality, we have adopted $\overline{u}=1.$

In the limit $z\rightarrow 0,$ the crossover function $Y\rightarrow 1,$ the
susceptibility and correlation length follow mean-field behavior $\chi
^{-1}=a_{0}\tau $ and $\xi =\overline{\xi }_{0}\tau ^{-1/2}.$ In the limit $%
z\rightarrow \infty ,$ $Y\rightarrow (\tau \xi _{{\rm D}}/\overline{\xi }%
_{0})^{\Delta _{{\rm S}}}\rightarrow 0,$ the susceptibility and the
correlation length exhibit Ising asymptotic critical behavior: $\chi =\Gamma
_{0}\tau ^{-\gamma }$ and\ $\xi =\xi _{0}\tau ^{-\nu }$ with amplitudes $%
\Gamma _{0}=0.871a_{0}^{-1}\left( \overline{\xi }_{0}/\xi _{{\rm D}}\right)
^{2(\gamma -1)}$ and $\xi _{0}=\overline{\xi }_{0}\left( \overline{\xi }%
_{0}/\xi _{{\rm D}}\right) ^{1-2\nu }$ \cite{AS:01}. Actually, the
Ornstein-Zernike approximation for the correlation function should replace
the Fisher-Burford\ approximation in the mean-field limit. However, the
difference between these two approximations is negligible when the
correlation length is small as it is in the mean-field regime.

To represent the experimental data for each molecular weight, four
parameters were used as adjustable: $I_{0},I_{{\rm b}},\overline{\xi }_{0},$
and $\xi _{{\rm D}}.$ Although the amplitude $a_{0}$ is absorbed in $I_{0}$,
to calculate $\Gamma _{0},$we fixed $a_{0}=1$, as predicted by the Flory
model in the {\it theta}-point limit\cite{PS:99}. The parameter $I_{{\rm b}}$
becomes important farther away from the critical temperature where $I_{{\rm b%
}}$\ and $\xi _{{\rm D}}$ are strongly statistically correlated. To make
sure that we found accurate values for the parameters $I_{{\rm b}}$\ and $%
\xi _{{\rm D}}$ we checked the results obtained for different fitting
intervals of $\ \tau .$

\begin{figure*}
\centering
\epsfbox{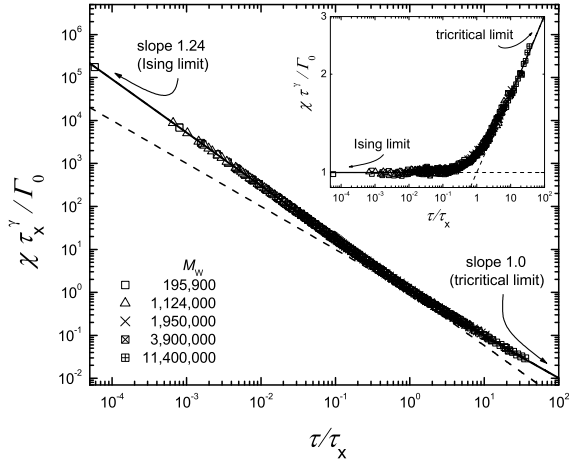}
\begin{minipage}[]{8.5cm}
\protect\caption{
Scaled osmotic susceptibility\ and deviation of the succeptibility
from Ising critical behavior (shown as insert) for solutions of polystyrene
in cyclohexane as a function of the scaled distance $\tau /\tau _{\times }$
to the critical temperature. The symbols represent experimental data, the
dashed lines represent two limiting behaviors: Ising asymptotic behavior and
mean-field behavior, while the solid curves represent the crossover theory.
}
\end{minipage}
\label{f:1}
\end{figure*}

\vspace*{-5mm}
\begin{figure*}
\centering
\epsfbox{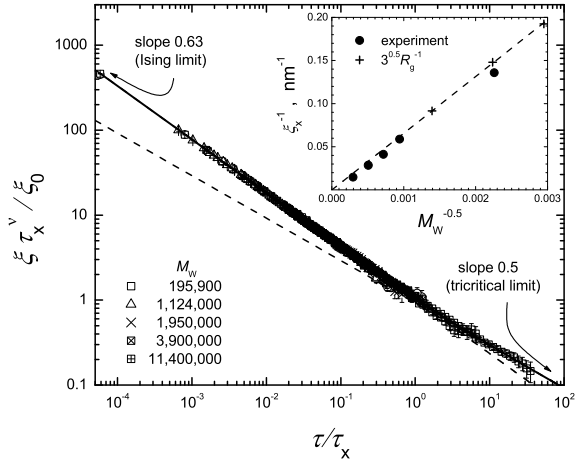}
\begin{minipage}[]{8.5cm}
\protect\caption{
Scaled correlation length for solutions of polystyrene in
cyclohexane as a function of the scaled distance $\tau /\tau _{\times }$ to
the critical temperature. The symbols represent experimental data, the
dashed lines represent Ising asymptotic behavior and mean-field behavior,
while the solid curve represents the crossover theory. In the insert:
molecular-weight dependence of the inverse correlation length at the
crossover temperature (crosses); closed circles are the normalized radius of
gyration, $R_{{\rm g}}/\sqrt{3\text{ }}$, scaled as squared root of the
molecular weight of polymer \protect\cite{MH:97}. 
For $M_{{\rm w}}\geq $ 10$^{6}$, \ the
values of $R_{{\rm g}\text{ }}$have been extrapolated.
}
\end{minipage}
\label{f:2}
\end{figure*}

A sensitive test of the shape of the crossover behavior can be obtained from
an analysis of the effective exponent of the susceptibility, defined as $%
\gamma _{{\rm eff}}=\partial \log \chi /\log \tau $. The exponent $\gamma _{%
{\rm eff}}$ exhibits crossover from its classical value $\gamma $ =1.00 to
its Ising value $\gamma $ =1.24. The reduced crossover temperature $\tau
_{\times }=(T_{\times }-T_{{\rm c}})/T_{c}$ can be defined as the inflection
point of $\gamma _{{\rm eff}}$. It turns out that $\tau _{\times }\cong (%
\overline{\xi }_{0}/\xi _{{\rm D}})^{2}.$

\begin{figure*}
\centering
\epsfbox{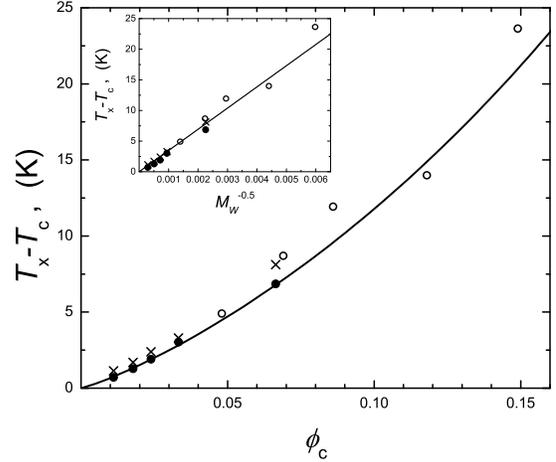}
\begin{minipage}[1pt]{8.5cm}
\protect\caption{
Difference between the crossover temperature $T_{\times }$ and the
critical temperature $T_{{\rm c}}$ of polystyrene-cyclohexane solutions as a
function of the critical volume fraction $\phi _{{\rm c}}$ and of $M_{{\rm w}%
}^{-1/2}$ (insert). Solid circles correspond to the inflection points of $%
\gamma _{{\rm eff}}(\tau ).$ Crosses correspond to the temperatures at which 
$\xi =R_{{\rm g}}/\sqrt{3}$. Open circles are the temperatures at which $\xi
=R_{{\rm g}}/\sqrt{3}$ for the system of polystyrene-deuterocyclohexane
\protect\cite{MH:97}. 
For $\phi _{{\rm c}}\leq 0.05$ the values of $R_{{\rm g}}$ have been
extrapolated. The solid curve is an approximation based on $(T_{\times }-T_{%
{\rm c}})\propto N^{-1/2}$, with $N(\phi _{c})$ defined parametrically by
Eq. (\ref{logcor}).
}
\end{minipage}
\label{f:3}
\end{figure*}

It follows from Eqs.(\ref{suscept}) and (\ref{corrlength}) that the
susceptibilities and the correlation lengths obtained for different
molecular weights (different $\xi _{{\rm D}})$ when reduced as $\chi \tau
_{\times }^{\gamma }/\Gamma _{0}$ and $\xi \tau _{\times }^{\nu }/\xi _{0},$
respectively, should collapse into master curves as functions of the scaled
temperature $\tau /\tau _{\times }.$ The master curve for the susceptibility
shown in Fig. 1 clearly demonstrates a crossover between two limits, Ising
and meanfield, over seven orders of $\tau /\tau _{\times }.$ In Fig. 2, the
same crossover is manifested by the correlation length. It is remarkable
that the correlation length taken at the crossover temperature follows
almost perfectly the normalized radius of gyration $R_{{\rm g}}/\sqrt{3}$,
independently measured by neutron scattering\cite{MW:97}. It also follows
from our analysis that the dependence of the critical amplitudes on the
molecular weight can be described within experimental accuracy by de Gennes'
scaling \cite{DG:79,W:93}. In Fig. 3, the difference between the crossover
temperature $T_{\times }$(determined as the inflection point of $\gamma _{%
{\rm eff}}(\tau )$ and as the temperature where $\xi =R_{{\rm g}}/\sqrt{3}$)
and the critical temperature is plotted as a function of the critical volume
fraction and of the molecular weight. The difference vanishes at the
{\it theta}-point limit and scales approximately as $M_{{\rm w}}^{-1/2},$ as
predicted by the crossover theory. In the Flory model, the critical volume
fraction $\phi _{{\rm c}}=1/(1+\sqrt{N})$, where $N=$ $M_{{\rm w}}/M_{0}$ is
the degree of polymerization ($M_{0}$ is the molecular weight of a
polymer-chain unit). According to de Gennes' scaling, $\phi _{{\rm c}}$ also
scales as $M_{{\rm w}}^{-1/2}$. However, the dependence of $T_{\times }-T_{%
{\rm c}}$ on $\phi _{{\rm c}}$ and, correspondingly, the dependence of $\phi
_{{\rm c}}$ on $\ M_{{\rm w}}^{-1/2}$ exhibit a pronounced non-linearity 
(Figs. 3 and 4). The violation of the Flory prediction for the critical
volume fraction is a well-known fact: it has been attributed to partial
collapsing of the polymer coils and described by a power law with an
additional critical exponent \cite{S:89,C:93}. Our analysis, which includes
old data of Melnichenko {\it et al. }\cite{MH:97} and our new data very
close to the {\it theta} point, 
show that such an exponent is not needed and that
all data are well described by the Flory-model formula corrected by a
renormalization-group-theory logarithmic term associated with tricritical
fluctuations:

\begin{equation}
\phi _{{\normalsize c}}=\frac{1}{1+\sqrt{N}(1+\upsilon \ln N)^{-1/2}}\text{ .%
}  \label{logcor}
\end{equation}
Eq.(\ref{logcor}) contains only one system-dependent parameter, $\upsilon
=1.26\pm 0.02,$ and, in contrast to the mean-field prediction, shows zero
slope at the {\it theta} point. Computer simulations reported recently, also
indicate the existence of logarithmic corrections \cite
{YP:00,G:97,FG:97,PF:98}. We suggest that the logarithmic correction is
indeed responsible for the observed behavior of $\phi _{c}$ as the new data
are much closer to the {\it theta} point and less affected by nonasymptotic
contributions.

\begin{figure*}
\centering
\epsfbox{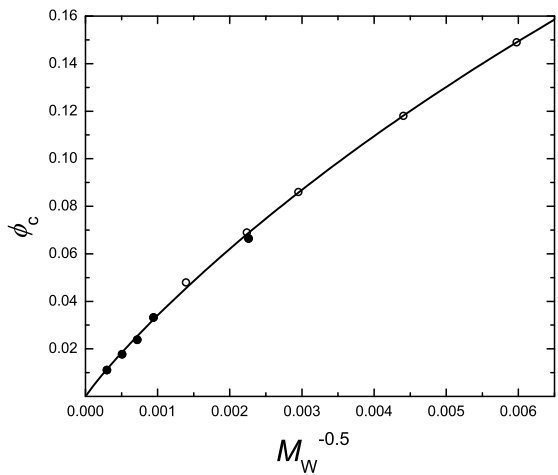}
\begin{minipage}[]{8.5cm}
\protect\caption{
Critical volume fraction $\phi _{{\rm c}\text{ }}$of solutions of
polystyrene in cyclohexane scaled as a function of the squared root of the
inverse molecular weight (closed circles). Open circles are the data for the
system of polystyrene-deuterocyclohexane \protect\cite{MH:97}. 
The solid curve represents Eq. (\ref{logcor}) with $\upsilon =1.26.$
}
\end{minipage}
\label{f:4}
\end{figure*}

We thank Y. G. Burya, V. A. Dechabo, R. W. Gammon, J. Jacob, and I. K. Yudin
for collaboration with various aspects of the experiment and to S. Wiegand
and M. Kleemeier for providing us with a Monte-Carlo simulation program. The
research has been supported by National Science Foundation Grant No.
CHE-9805260.

\end{multicols}
\vfill

\end{document}